\documentclass[onecolumn,journal,12pt,draftclsnofoot]{IEEEtran}

\usepackage{amsmath,amsthm,bm} 
\usepackage{units}
\IEEEoverridecommandlockouts
\usepackage{graphicx}

\usepackage[cmintegrals]{newtxmath}
\usepackage{mathtools}

\usepackage[short]{optidef}

\usepackage{afterpage}
\usepackage{verbatim}
\usepackage{psfrag}
\usepackage{color}
\usepackage[table]{xcolor}
\usepackage{cite}
\usepackage{times}
\usepackage{wrapfig}
\usepackage{pgfplots}
\usepackage{multirow}
\usepackage{tikz}
\usepackage{setspace}
\usepackage{epsfig}
\usepackage{epstopdf}
\usepackage{footmisc}
\usepackage{fmtcount}
\usepackage{stackrel}
\usepackage{float}
\usepackage{enumerate}
\usepackage{subcaption}
\usepackage{hyperref}
\usepackage{hypcap}
\usepackage[nottoc]{tocbibind}
\usepackage[normalem]{ulem}
\usepackage{authblk}
\usepackage{tabularx}
\usepackage{bbm}
\usepackage{array}
\usepackage{listings}

\DeclarePairedDelimiter\floor{\lfloor}{\rfloor}
\usepackage{algorithm}
\usepackage{algpseudocode}

\usepackage{stackengine}
\setstackEOL{\\}
\usepackage{tabularx}
\makeatletter

\usepackage[T1]{fontenc}
\usepackage[utf8]{inputenc}
\usepackage{babel}
\usepackage[labelfont=bf]{caption}

\newcommand{\LinesNumbered}{%
  \setboolean{algocf@linesnumbered}{true}%
  \renewcommand{\algocf@linesnumbered}{\everypar={\nl}}}%
  
\let\oldnl\nl
\newcommand{\nonl}{\renewcommand{\nl}{\let\nl\oldnl}}

\newcommand{\multiline}[1]{%
  \begin{tabularx}{\dimexpr\linewidth-\ALG@thistlm}[t]{@{}X@{}}
    #1
  \end{tabularx}
}
\makeatother

\usepackage{capt-of}

\begin{document}
\bstctlcite{IEEEexample:BSTcontrol}

\title{Deep Reinforcement Learning for Optimizing RIS-Assisted HD-FD Wireless Systems}

\author{Alice~Faisal,~\IEEEmembership{Student Member,~IEEE,}
        Ibrahim~Al-Nahhal,~\IEEEmembership{Member,~IEEE,}
        Octavia~A.~Dobre,~\IEEEmembership{Fellow,~IEEE}
        and~Telex~M. N.~Ngatched,~\IEEEmembership{Senior Member,~IEEE}
\thanks{
A. Faisal, I. Al-Nahhal, O. A. Dobre, and T. M. N. Ngatched are with the Faculty of Engineering and Applied Science, Memorial University, St. John’s, NL,
Canada, (e-mail: {afaisal, ioalnahhal, odobre}@mun.ca; tngatched@grenfell.mun.ca).
}} %

\maketitle
\begin{abstract}
This letter investigates the reconfigurable intelligent surface (RIS)-assisted multiple-input single-output (MISO) wireless system, where both half-duplex (HD) and full-duplex (FD) operating modes are considered together, for the first time {in the literature}. The goal is to maximize the rate by optimizing the RIS phase shifts. {A} novel deep
reinforcement learning (DRL) algorithm is proposed to solve the {formulated} non-convex optimization problem. The complexity analysis and Monte Carlo simulations illustrate that the proposed DRL algorithm significantly improves the rate compared to the non-optimized scenario in both HD and FD operating modes using  {a single parameter} {setting.} Besides, it significantly reduces the  computational complexity of the downlink HD MISO system and improves the achievable rate with a {reduced} number of steps per episode compared to the {conventional} DRL algorithm. 
\end{abstract}
\begin{IEEEkeywords}
Reconfigurable intelligent surface (RIS), {half-duplex full-duplex (HD-FD)}, deep reinforcement learning (DRL).
\end{IEEEkeywords}

\IEEEpeerreviewmaketitle
\section{Introduction}
\IEEEPARstart{R}{econfigurable}  {intelligent surfaces}  {(RIS)} have emerged as a promising paradigm to fulfill  {the need of a smart and programmable wireless environment, and meet the demands of future wireless networks \cite{9350614,RIS_O}.} RIS consists of a two-dimensional array of low-cost passive electromagnetic (EM) elements \cite{Access}. By overcoming the random nature of EM wave propagation, RIS enables controlling different characteristics of radio waves, such as scattering, reflection, and refraction. Consequently, it  {effectively enhances the} signal quality and boosts the wireless spectral efficiency by realizing a controllable environment \cite{RIS_Survay2}.

RIS-assisted multiple-input multiple-output systems have  {recently} drawn significant attention as a cost-effective solution to enhance the wireless transmission in both half-duplex (HD)  {and} full-duplex (FD) operating modes \cite{HD_Beam,HD_robust,HD_rate,shen2020beamforming,EE-FD,multiuser-FD}. In the HD mode, systems require additional resources
to receive and forward signals, which results in a decreased spectral efficiency. In contrast, the FD mode has the potential to significantly increase the throughput of wireless systems as it enables  {simultaneous transmission and reception of signals} in the same frequency band. However, this comes at the cost of {increased interference} and implementation complexity. To this end, some researchers are considering HD-FD transmission schemes {that} combine the advantages of both HD and FD modes \cite{elhattab2021reconfigurable}. In \cite{HD_Beam} and \cite{HD_robust}, RIS-HD systems are optimized to minimize the total transmit power. In \cite{HD_rate}, a joint optimization problem is considered to maximize the achievable rate of {an} RIS-HD system. In \cite{shen2020beamforming} and \cite{EE-FD}, the sum-rate and spectral efficiency of {an} RIS-FD system is maximized, respectively. In \cite{multiuser-FD}, the  weighted minimum rate is maximized for a multi-user {RIS-FD} system.
Most of these works decoupled the optimization variables using alternating optimization algorithms, which {exhibit} both loss of optimality and high computational complexity.

Deep learning  {has emerged} as a powerful approach to optimize the RIS {phase shifts} by tackling the practical implementation problems of the optimization techniques \cite{dl_sur,DL_layering}. In particular,  {deep reinforcement learning} (DRL) {is a potential candidate} to optimize the RIS phase shifts without  {the} need for offline training with a labeled dataset.
 {A few} works have considered DRL approaches to optimize  RIS-HD systems \cite{DL_robust,huang2020reconfigurable, DRLmain}. {The authors in \cite{DL_robust} proposed an optimization-driven deep deterministic policy gradient (DDPG) to minimize the access point's transmit power. The sum-rate maximization problem of {a} multi-user RIS-HD system was addressed in \cite{huang2020reconfigurable} using a DRL algorithm.}
 Furthermore, a {conventional} DRL algorithm is introduced in \cite{DRLmain} to maximize the received signal-to-noise ratio of the downlink RIS-HD {multiple-input single-output (MISO)} system. To the best of the authors’ knowledge, utilizing DRL for RIS-FD systems has  {not yet been} discussed in the literature.

In this letter, a novel DRL algorithm is proposed to optimize the phase shifts of {an} RIS-assisted HD-FD MISO system. The contributions are summarized as follows:

\begin{itemize}
    \item 
    The proposed DRL algorithm achieves promising results in the HD and FD {operating modes} without  {the} need of additional parameters tuning. 
    \item It provides a significant {improvement in the} rate compared to the non-optimized RIS phase shifts in the HD and FD operating modes.
    \item It significantly reduces the  computational complexity, while providing a considerable rate improvement with a  {reduced} number of required steps {for each} episode, compared to the {conventional} DRL in \cite{DRLmain} {for} the HD mode.
    \item {The} complexity analysis and Monte Carlo simulations {support} the {findings}.

\end{itemize}

The  {remainder} of this  {letter} is organized as follows: Section \ref{system_model} presents the system model and problem formulation for the  {RIS-assisted HD-FD} MISO system. The proposed DRL algorithm is  {introduced} in Section \ref{solution}, {and} its computational complexity  {is analyzed in Section \ref{Complexity}}. {Simulation} results and conclusions are presented in Sections \ref{simulations} and \ref{conclusion}, respectively.



\section{System Model and Problem Formulation} \label{system_model}

\begin{figure}[t]
  \centering
  \includegraphics[scale = 0.40]{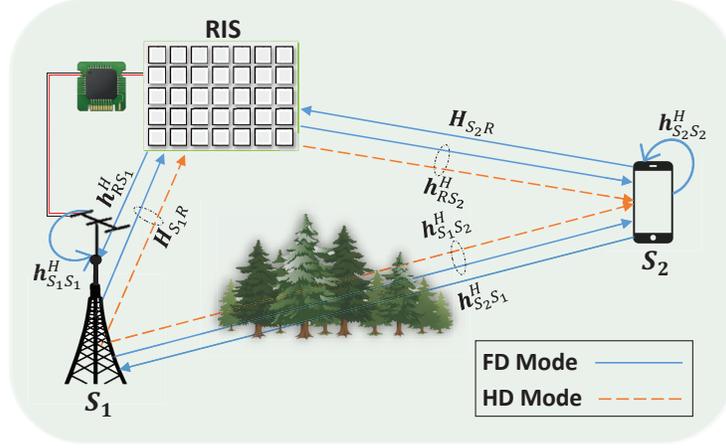}
  \caption{ {RIS-assisted HD-FD} MISO system. 
  }
  \label{fig:HD system}
\end{figure}

Consider an  {RIS-assisted HD-FD} MISO system as illustrated in Fig. \ref{fig:HD system}, where $S_1$ and $S_2$ represent the  {base station (BS)} and user equipment (UE), respectively.  {Both the BS and UE} are equipped with $M$ transmit antennas and one receive antenna. The UE sometimes operates in a HD mode, where it only receives information from the BS (i.e., downlink HD mode), {while other} times the UE and BS transmit and receive information simultaneously in the same frequency band (i.e., FD mode).  {Henceforth, $\Omega$ denotes} the operating mode, where $\Omega \in \{\text{HD}, \text{FD}\}$. The RIS is composed of $N$ programmable reflecting elements, which assists the communication between $S_1$ and $S_2$ by optimizing the RIS phase shifts through an RIS controller. Given $\bar{i} = 3 - i$  {$\forall \hspace{0.2em}i = 1,2$}, let $\mathbf{H}_{{S_{\bar{i}}R}}\in\mathbb{C}^{N\times M}$, $\mathbf{h}^H_{RS_i}\in\mathbb{C}^{1\times N}$, and $\mathbf{h}^H_{{S_{\bar{i}} S_{i}}}\in\mathbb{C}^{1\times M}$
denote the channel coefficients of {the} $S_{\bar{i}}$-RIS, RIS-$S_i$, and $S_{\bar{i}}$-$S_i$ links, respectively. {The self-interference (SI) channels, which are involved in the FD mode at the BS and  {UE are} denoted by $\mathbf{h}^H_{S_iS_i} \in\mathbb{C}^{1\times M}$.}

At the  {receiver-side}, the signal is received from the direct and reflected links of the BS and RIS, respectively. Thus, the {noisy} received {signals} of the downlink HD and FD operating modes are respectively expressed as
\begin{equation}
    y_i^{\Omega} = \Bigl(\underbrace{\mathbf{h}_{RS_i}^H\boldsymbol{\Theta}\mathbf{H}_{S_{\bar{i}}R}}_{\text{{Reflected signal}}} \hspace{2mm} + \hspace{-1mm} \underbrace{\mathbf{h}_{{S_{\bar{i}} S_{i}}}^H}_{\text{Direct signal}}\hspace{-2mm}\Bigr)\mathbf{w}_{\bar{i}}x_{\bar{i}} + n, \hspace{0.1em} i = 2, \hspace{0.1em} \Omega = \text{HD},
    \end{equation}
    \noindent and
\begin{equation}
    y_i^{\Omega} =\Bigl(\underbrace{\mathbf{h}_{RS_i}^H\boldsymbol{\Theta}\mathbf{H}_{S_{\bar{i}}R}}_{\text{Reflected signal}} \hspace{0mm} + \hspace{-3mm} \underbrace{\mathbf{h}_{{S_{\bar{i}} S_{i}}}^H}_{\text{\hspace{1mm}Direct signal}}\hspace{-3mm}\Bigr)\mathbf{w}_{\bar{i}}x_{\bar{i}} \hspace{0.1em} + \hspace{0.1em} \underbrace{\mathbf{h}_{{S_{i} S_{i}}}^H \mathbf{w}_ix_i}_{\text{Residual SI}} + \hspace{0.1em} n, \\ \hspace{0.1em} i = 1, 2, \Omega = \text{FD},
    \label{FD_received}
    \end{equation}

\noindent where $n\sim\mathcal{CN}(0,\sigma^2)$   {denotes} the additive white {complex} Gaussian noise {with zero-mean and {variance $\sigma^2$}.} The diagonal matrix $\boldsymbol{\Theta} = \text{diag}\bigl(e^{j\varphi_1}, \cdots,e^{j\varphi_n},\cdots,e^{j\varphi_N}\bigr)\in\mathbb{C}^{N\times N}$  {represents} the phase shifts of the RIS,  {where}  $\varphi_n\in[-\pi,\pi)$  {is the phase shift introduced} by the $n$-th reflecting {element.} 
The source node, $S_i$, employs an active beamforming $\mathbf{w}_i\in\mathbb{C}^{M\times 1}$ to transmit the information signal, $x_i$, with $\mathbb{E}\{|x_i|^2\}=1$, where $\mathbb{E}\{\cdot\}$ denotes the expectation operation. {The third term in \eqref{FD_received} represents the SI introduced by the FD mode operation.} 

{The} achievable rate and sum-rate of the downlink HD and FD operating {modes, measured in bit per second per Hertz (bps/Hz), are respectively given as}

\begin{equation}
    \mathcal{R}^{\Omega} = \text{log}_2\left(1+ \frac{\left|\left(\mathbf{h}_{RS_i}^H\boldsymbol{\Theta}\mathbf{H}_{S_{\bar{i}}R} +  \mathbf{h}_{{S_{\bar{i}} S_{i}}}^H \right)\mathbf{w}_{\bar{i}}\right|^2}{\sigma^2}\right), \hspace{0.1em} i = 2, \hspace{0.1em} \Omega = \text{HD},
    \label{HD:reward}
 \end{equation} 
    \noindent and
  \begin{equation}  
    \mathcal{R}^{\Omega} = \sum_{i=1}^2 \text{log}_2\left(1+ \frac{\left|\left(\mathbf{h}^H_{RS_{{i}}} \boldsymbol{\Theta}\mathbf{H}_{S_{\bar{i}}R} + \mathbf{h}^H_{S_{\bar{i}}S_i}\right)\mathbf{w}_{\bar{i}}\right|^2}{|\mathbf{h}^H_{S_iS_i} \mathbf{w}_i|^2 + \sigma^2}\right), \hspace{0.1em} \Omega = \text{FD}.
    \label{FD:reward}
\end{equation}

\noindent {Here}, the goal is to maximize the rate of the  {RIS-assisted HD-FD} {MISO} system by optimizing the RIS phase shifts. Thus, the resulting optimization problem can be expressed as


\begin{maxi!}
{ \boldsymbol{\varphi}}{\mathcal{R}^{\Omega}, \hspace{1mm} \Omega \in \{\text{HD},\text{FD}\} } 
{}{\text{(P1)}\quad}
\addConstraint{{ -\pi \leq \varphi_n \leq \pi, \hspace{0.2em} n = 1, \cdots , N.}} \label{eq:IntRIS42}
\end{maxi!}





\noindent It is worth noting that the {conventional} DRL algorithm in \cite{DRLmain} has been proposed to solve the non-convex problem (P1) only when $\Omega$ = HD,  {and} suffers from  high computational complexity. Moreover, the DRL for the FD operating mode has  {not yet been} investigated  {in the literature}.

\section{Proposed DRL Algorithm}
\label{solution}
This section proposes a novel DRL algorithm to solve (P1) for the  {RIS-assisted HD-FD} MISO system. To deal with (P1), the RIS phase shifts are optimized using the proposed DRL algorithm. Then, for a given optimized $\boldsymbol{\Theta}$, the transmit beamformers, $\mathbf{w}_{\bar{i}}$, are optimized using a closed and semi-closed form solutions for the HD and FD operating modes, respectively. The optimization problem is solved in an iterative fashion until the optimized $\boldsymbol{\Theta}$ and $\mathbf{w}_{\bar{i}}$ converge.


\subsection{Beamforming Design for a Given $\mathbf{\Theta}$}

The optimal beamforming vector for the HD operating mode is calculated using the maximum ratio transmission approach, whereas a semi-closed optimal solution of the FD beamforming vectors is given in \cite{shen2020beamforming}. Consequently, for a given optimized $\boldsymbol{\Theta}$, the optimal beamforming vectors of the HD and FD modes, $\mathbf{w}_{\bar{i}}$, are respectively given as
\begin{equation}
\mathbf{w}_{\bar{i}}^\dagger = \sqrt{P_{\text{max}}} \frac{\left(\mathbf{h}_{RS_i}^H\boldsymbol{\Theta}\mathbf{H}_{S_{\bar{i}}R} +  \mathbf{h}_{{S_{\bar{i}} S_{i}}}^H \right)^H}{\left|\left|\left(\mathbf{h}_{RS_i}^H\boldsymbol{\Theta}\mathbf{H}_{S_{\bar{i}}R} +  \mathbf{h}_{{S_{\bar{i}} S_{i}}}^H \right)\right|\right|}, \hspace{0.1em} i = 2, \hspace{0.1em} \Omega = \text{HD},
\label{beamformingHD}
\end{equation}
\noindent and
\begin{equation}
    \mathbf{w}^\dagger_{\bar{i}} = (\delta\mathbf{h}_{S_{\bar{i}} S_{\bar{i}}} \mathbf{h}^H_{S_{\bar{i}} S_{\bar{i}}} + v^\dagger \mathbf{I})^{-1} \boldsymbol{\mathcal{B}}, \hspace{0.1em}  i = 1, 2, \hspace{0.1em} \Omega = \text{FD},
    \label{beamformingFD}
\end{equation}


\noindent  where $P_{\text{max}}$ is the maximum transmitted power of $S_{\bar{i}}$, $\mathbf{I}$ is the identity matrix, and $v^\dagger$ is the optimal dual {Lagrangian} variable associated with the power constraint that is found by performing a bisection search over the interval $\left[0, \sqrt{
\boldsymbol{\mathcal{B}}^T \boldsymbol{\mathcal{B}}}/\sqrt{P_{\text{max}}}\right]$. Here, $\boldsymbol{\mathcal{B}}$ and $\delta$ are given as


\begin{equation}
 \boldsymbol{\mathcal{B}} \triangleq \frac{1}{\tilde{b}_{i}} \left(1+ \frac{b_{{i}}}{|\mathbf{h}_{S_{\bar{i}} S_{\bar{i}}}^H \mathbf{\tilde{w}}_{\bar{i}}|^2+\sigma^2}\right) \mathbf{h}_{\bar{i}}\mathbf{h}^H_{\bar{i}} \mathbf{\tilde{w}}_{\bar{i}}, 
         \end{equation}
         \noindent and
        \begin{equation}
    \delta \triangleq \frac{b_{{i}}\left(|\mathbf{h}^H_{\bar{i}} \mathbf{\tilde{w}}_{\bar{i}}|^2+\tilde{b}_{i}\right)}{\tilde{b}_{i}\left(|\mathbf{h}_{S_{\bar{i}} S_{\bar{i}}}^H \mathbf{\tilde{w}}_{\bar{i}}|^2+\sigma^2\right)^2},
\end{equation}

 \noindent where $b_{{i}} \triangleq |\mathbf{h}^H_{{i}} \mathbf{w}_{{i}}|^2$, $\tilde{b}_{i} \triangleq |\mathbf{h}^H_{S_{i}S_{{i}}}\mathbf{w}_{i}|^2$  {$+$} $\sigma^2$, $\mathbf{h}_{\bar{i}} \triangleq \mathbf{H}^H_{S_{\bar{i}}R}\boldsymbol{\Theta}^H\mathbf{h}_{RS_{{i}}} + \mathbf{h}_{S_{\bar{i}}S_{{i}}}$, and $\mathbf{\tilde{w}}_{\bar{i}}$ is a given
feasible point.


\subsection{Phase  {Shift} Design Based on the Proposed DRL Algorithm}

\subsubsection{Problem Transformation}

The RIS controller represents the DRL \textit{agent}, while the  {RIS-assisted HD-FD} MISO communication system represents the DRL \textit{environment}. Thus, the \textit{state space}, \textit{action space}, and \textit{reward} for the proposed DRL algorithm are defined as follows:

\begin{itemize}
    \item State space: The state space at time step $t$, {$s_t \in \mathbb{R}^{1 \times (N+1)}$}, includes $\varphi_n \hspace{0.1em}  {\forall \hspace{0.2em} n} = 1, \cdots, N$ and the corresponding $\mathcal{R}^{\Omega}$  at time step $t-1$, {and is defined as} 
    \begin{equation}
    s_t = \left[\mathcal{R}^{\Omega, (t-1)}, \varphi^{(t-1)}_1, \cdots,\varphi^{(t-1)}_n,\cdots,\varphi^{(t-1)}_N\right].
    \end{equation}
    
    \item Action space: Since (P1) aims to optimize the RIS phase shifts, the action space at time step $t$, {$a_t \in \mathbb{R}^{1 \times N}$}, is expressed as 
    \begin{equation}
        a_t = \left[\varphi^{(t)}_1, \cdots,\varphi^{(t)}_n,\cdots,\varphi^{(t)}_N\right].
        \end{equation}
    
    \item Reward: As the target of (P1) is to maximize $\mathcal{R}^{\Omega}$, the reward is expressed as \begin{equation}
        r_t = \mathcal{R}^{\Omega, (t)}, \hspace{1mm} \Omega \in \{\text{HD},\text{FD}\}. 
        \end{equation}
\end{itemize}

At each time step $t$, the agent receives the current state $s_t$ from the environment, takes an action $a_t$ based on  {a} \textit{policy} $\tilde{\pi}$, and receives a scalar reward $r_t$. Then, a new state $s_{t+1}$ is obtained. The return of a state is defined as the \textit{total discounted reward} from time step $t$ onwards, {and is given by} $R_t = \sum_{k=t}^\infty \gamma^{k-t} r(s_k,a_k)$, where $\gamma \in (0,1]$ is the DRL discount factor. The goal is to learn a policy that maximizes the expected cumulative discounted reward from the start state, as: $J(\tilde{\pi}) = \mathbb{E}\left[R_1 | \tilde{\pi}\right]$. 
{The {DDPG}, which combines the benefits of value-based and policy-based approaches \cite{lillicrap2019continuous}, is used to learn the optimal policy for a continuous $a_t$.} In particular, the DDPG algorithm aims at maximizing
the Q-value of $(s,a)$ pair by training a deep neural network (DNN), defined as

\begin{equation}
    Q^{\tilde{\pi}_{\boldsymbol{\theta}}}(s,a) = \mathbb{E}_{\tilde{\pi}_{\boldsymbol{\theta}}} \biggl[R_1 |s_1 = s, a_1 = a\biggr],
\end{equation}

\noindent where $\boldsymbol{\theta}$  {represents} the DNN parameters, as well as finding the optimal policy by performing  {the} gradient ascent of
\begin{equation}
\nabla_{\boldsymbol{\theta}} J(\tilde{\pi}_{\boldsymbol{\theta}}) = \mathbb{E}_{\tilde{\pi}_{\boldsymbol{\theta}}}\left[Q^{\tilde{\pi}_{\boldsymbol{\theta}}}(s,a)\nabla_{\boldsymbol{\theta}} \text{log}\pi_{\boldsymbol{\theta}}(a|s) \right].
\end{equation}

\begin{figure}[t]
  \centering
  \includegraphics[scale=0.4]{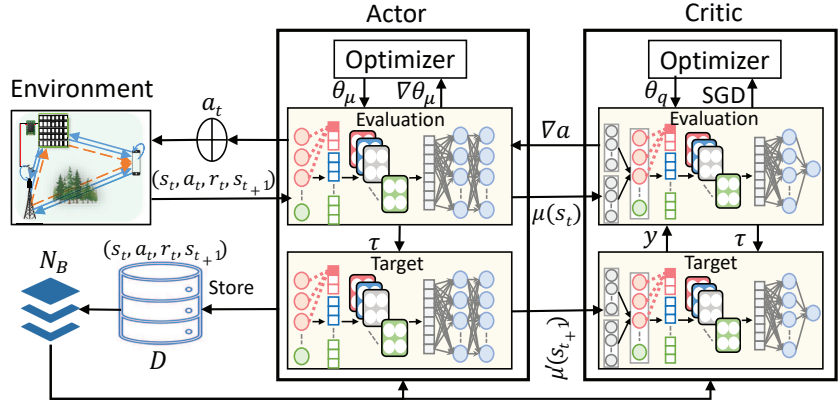}
  \caption{The proposed DRL algorithm structure.}
  \label{fig:DDPG}
\end{figure} 

 {The DDPG algorithm} is based on the actor-critic technique, which consists of two DNN models: actor and critic. The actor, $\mu(s_t|\boldsymbol{\theta}_\mu)$, represents the policy network that takes the state as an input for a given $\boldsymbol{\theta}_\mu$ and outputs  $a_t = \mu(s_t|\boldsymbol{\theta}_\mu) + \xi$, where $\xi$ is a random process that is added to the actions for exploration. {$\xi$ is modeled as complex Gaussian process with zero mean and variance 0.1.} The critic, $Q(s_t,a_t|\boldsymbol{\theta}_q)$, represents the network that evaluates the  {actions. It} takes $s_t$ and $a_t$ as an input for a given $\boldsymbol{\theta}_q$, and outputs the Q-value. The  {DDPG algorithm} utilizes the concept of experience replay with memory $D$ to reduce the correlation of the training samples by randomly sampling minibatch transitions, $N_B$. Moreover, target networks are introduced to stabilize the learning process. The target networks are generated by making a copy of the actor and critic evaluation NNs, $\mu'(s_t|\boldsymbol{\theta}_{\mu'})$ and $Q'(s_t,a_t|\boldsymbol{\theta}_{q'})$, and are used to calculate the corresponding target values, $y_t$ in \eqref{target eq}. The  {actor and critic NN parameters, $\boldsymbol{\theta}_\mu$ and $\boldsymbol{\theta}_q$}, are updated using the stochastic gradient descent (SGD) from \eqref{SGD eq} and policy gradient from \eqref{pg eq}, respectively. Finally, the target NN parameters are updated using a soft update coefficient, $\tau$, based on \eqref{taR1 eq} and \eqref{TAR2 eq}. After $T$ steps of each episode, the {agent's} performance saturates and  {it} outputs the optimized $\Theta$. The structure of the proposed DRL algorithm is illustrated in Fig. \ref{fig:DDPG} and summarized in Algorithm 1.

\begin{algorithm}[h!]
\caption{Proposed DRL  {algorithm.}}
\begin{algorithmic}[1]
\Require{$\boldsymbol{\theta}_\mu$ and $\boldsymbol{\theta}_q$ with random weights, $D$, $\gamma$, $\tau$, and learning rate $\alpha$;

\hspace{-17mm} \textbf{Set:} $\boldsymbol{\theta}_{\mu'} \leftarrow \boldsymbol{\theta}_\mu$ and $\boldsymbol{\theta}_{q'} \leftarrow \boldsymbol{\theta}_q$;}

\Repeat 
\State \multiline{Collect the channels of the $k$-th episode based on $\Omega$;}
\State \multiline{Randomly initialize $\varphi_n \hspace{0.2em}  {\forall \hspace{0.2em}} n = 1, \cdots, N$ to obtain the initial state;}

\If{$\Omega = \text{HD}$}
\State{Calculate $\mathbf{w}_{\bar{i}}$ using \eqref{beamformingHD}};
\Else
\State{Calculate  {$\mathbf{w}_{\bar{i}}$} using \eqref{beamformingFD}};
\EndIf

\State Initialize  $\xi\sim\mathcal{CN}(0,0.1)$;
\Repeat 
\State \multiline{Obtain $a_t = \mu(s_t|\boldsymbol{\theta}_\mu) + \xi$ from the actor network and reshape it;}
\State{Repeat \textbf{Lines} \#4-8;}
\State{Observe the new state, $s_{t+1}$, given $a_t$; }
\State \multiline{Store ($s_t,a_t,r_{t},s_{t+1}$) in $D$;}
\State  {\multiline{When $D$ is full, sample a minibatch of $N_B$ transitions randomly ($s_j,a_j,r_{j},s_{j+1}$) from $D$;}}
\State{Compute the target value using target networks: 
\begin{equation}
    y_j = r_j + \gamma Q'(s_{j+1},\mu'(s_{j+1}|\boldsymbol{\theta}_{\mu'})|\boldsymbol{\theta}_{q'});
    \label{target eq}
\end{equation}}

\State \multiline{Update  {the critic} by minimizing the loss using SGD:
\vspace{-3mm}
\begin{equation}
    L = \frac{1}{N_B} \sum_j \left(y_j - Q(s_j,a_j|\boldsymbol{\theta}_q)\right)^2;
    \label{SGD eq}
\end{equation}}
\State Update the actor using the policy gradient: 
\begin{equation}
    \nabla_{\boldsymbol{\theta}_\mu} = \frac{1}{N_B}\sum_j \nabla_a Q(s,a|\boldsymbol{\theta}_q)|_{s=s_j,a = \mu(s_j)} \nabla_{\boldsymbol{\theta}_\mu} \mu(s|\boldsymbol{\theta}_\mu)|_{s_j};
    \label{pg eq}
\end{equation}
\State{Update the target NNs through soft update: 
\begin{equation}
    \boldsymbol{\theta}_{q'} \longleftarrow \tau \boldsymbol{\theta}_{q} + (1-\tau) \boldsymbol{\theta}_{q'},
    \label{taR1 eq}
\end{equation}
\begin{equation}
    \boldsymbol{\theta}_{\mu'} \longleftarrow \tau \boldsymbol{\theta}_{\mu'} + (1-\tau) \boldsymbol{\theta}_{\mu'}. 
    \label{TAR2 eq}
\end{equation} 
 \Until{$t = T$;} 
 \Until{$k = K$;} }
\Ensure{Optimal action that corresponds to the optimal $\boldsymbol{\Theta}$.}
\end{algorithmic}
\label{al:BCD}
\end{algorithm}

\subsubsection{Proposed  DNN Design} \label{DNN:sec}
As can be seen from Fig. \ref{fig:DDPG}, the proposed DRL algorithm contains four NNs (i.e., two NNs for  {the actor and two NNs for the critic).} A novel design is proposed for the four NNs, which consists of  {the} input layer, two hidden layers and  {the} output layer. The two hidden layers are a combination of one convolutional layer and one feed-forward (FF) layer with a flatten layer between them. The input layer of the actor and critic networks contains $N + 1$ neurons (i.e., size of $s_t$) and $2N + 1$ neurons (i.e., concatenation of $s_t$ and $a_t$), respectively. The output layer of the actor and critic networks contains $N$ neurons (i.e., size of $a_t$) and one neuron (i.e., scalar Q-value), respectively. {The convolutional hidden layer for each of {the} actor and critic networks {uses the} \textit{ReLU} activation function since it does not suffer from vanishing or exploding problems. In contrast, the FF hidden layer {uses} the \textit{softmax} activation function to obtain probabilistic values for all inputs.}

\section{Complexity Analysis} \label{Complexity}
The computational complexity of the {conventional} DRL algorithm in \cite{DRLmain} and  {the} proposed DRL algorithm for {$\Omega = \text{HD}$} is derived in terms of the number of NN parameters $C_\mathcal{P}$ required to be stored, real additions $C_\mathcal{A}$, and real multiplications $C_\mathcal{M}$. 
The  {conventional} DRL algorithm uses two hidden FF layers,  {and} its computational complexity is given as
\begin{align}
    & C_\mathcal{P} = \sum_{i=1}^3 (\eta_{i}+1)\eta_{i+1}, \\
    & C_\mathcal{M} = \sum_{i=1}^3 \eta_i \eta_{i+1}, \\
    & C_\mathcal{A} = \sum_{i=1}^3 \eta_i \eta_{i+1} + \sum_{i=1}^3 \eta_{i+1},
\end{align}

\noindent where $\eta_i$ is the number of neurons of the $i$-th layer. It is worth noting  {that, for simplicity, each} activation function is considered to cost one real addition.


Based on the NNs design in Section \ref{DNN:sec}, the complexity for the proposed DRL algorithm is given as
\begin{align}
       & C_\mathcal{P} = (\eta_F \eta_3+F_z +1)F_n + (\eta_4 +1)\eta_3 + \eta_4, \\
    & C_\mathcal{M} = (F_z+\eta_3)\eta_F F_n +\eta_3 \eta_4, \\
    & C_\mathcal{A} = (F_z+\eta_3+1)\eta_F F_n +(\eta_4+1)\eta_3 + \eta_4,
\end{align}

\noindent where $\eta_F = \floor{\frac{\eta_1 -  {F_z}}{F_s}+1}$,  {with} $\floor{\cdot}$  {as} the floor operation, $F_z$ is the filter size, $F_n$ is the number of filters, and $F_s$ is the stride. The complexity reduction of using the proposed DRL algorithm over the  {conventional} one for $\Omega = \text{HD}$ is
\begin{equation}
    \text{Reduction} =  1 - \frac{ \left\{C^{\text{Actor}}_\chi + C^{\text{Critic}}_\chi\right\}_{\text{Proposed}}}{\left\{C^{\text{Actor}}_\chi + C^{\text{Critic}}_\chi \right\}_{\text{{Conventional}}}}, \hspace{0.2em} \chi \in \{\mathcal{P}, \mathcal{A}, \mathcal{M}\}.
\end{equation}

\section{Simulation Results}\label{simulations}

\begin{figure}[t]
    \centering
    \includegraphics[scale = 0.4]{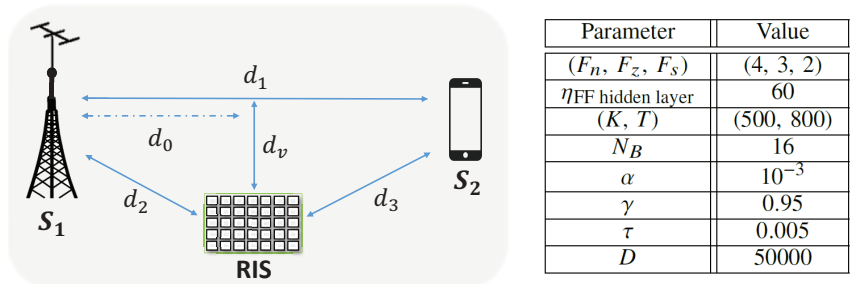}
    \caption{Simulation setup and DDPG parameters.}
    \label{fig:setup}
\end{figure}

This section evaluates the performance of the proposed DRL algorithm for the  {RIS-assisted HD-FD} MISO system. The simulation setup is shown in Fig. \ref{fig:setup}, where the considered parameters are $d_v = \unit[2]{m}$ and $d_1 = \unit[50]{m}$. The  {distances} of the BS-RIS and UE-RIS links  {are} calculated as $d_2 = \sqrt{d^2_0+d^2_v}$ m and $d_3 = \sqrt{(d_1-d_0)^2+d^2_v}$ m,  {respectively}. The path loss (PL) at distance $d_j$, $  \forall j \in \{1,2,3\}$ is modeled as $\text{PL} = PL_0 - 10 \zeta \text{log}_{10}\left(\frac{d_j}{ {D_r}}\right)$ \cite{DRLmain}, where $PL_0$ is the PL at a reference distance $ {D_r}$ and $\zeta$ is the PL exponent, in which $PL_0 =  {\unit[-30]{dB}}$ and $ {D_r} = \unit[1]{m}$.
 {As in \cite{DRLmain}}, the {BS-UE} channels are modeled as Rayleigh fading  {(assuming a blocking element between $S_1$ and $S_2$)}, while the rest of the channels are Rician with a factor of $10$. The PL exponents of  {the} BS-UE, BS-RIS, and UE-RIS channels are set to $\zeta_{\text{BU}} = 3$ and $\zeta_{\text{BR}} =  \zeta_{\text{UR}} = 2$, respectively. The PL of the SI channels for the FD mode is $\unit[-95]{dB}$. The total transmit power is $P = \unit[5]{dBm}$, while the noise power is $\sigma^2 = \unit[-80]{dBm}$. The antenna gain at the BS and UE is $\unit[0]{dBi}$, while the RIS gain is $\unit[5]{dBi}$. The penetration loss in  {the} BS-UE and RIS-UE links is $\unit[10]{dB}$.


 {The parameters of the proposed DRL algorithm are summarized in Fig. \ref{fig:setup}. Furthermore, the design {of the NNs} is explained in Section \ref{DNN:sec}, {and} its parameters are provided in Fig. \ref{fig:setup}. {The} Adam optimizer {is used to} update the parameters {of the NNs}.}
To {assess} the performance of the proposed algorithm, it is compared with the non-optimized scenario, referred to as random phase shifts. The  {conventional} DRL algorithm in \cite{DRLmain} with $T = 1000$ is {also} included to show the superiority of the proposed DRL algorithm in the HD mode.  {It is worth noting that the current form of the  {conventional} DRL algorithm can not be used to optimize the RIS phase shifts in the FD mode.}

\begin{figure}[h]
    \centering
    \includegraphics[scale=0.4]{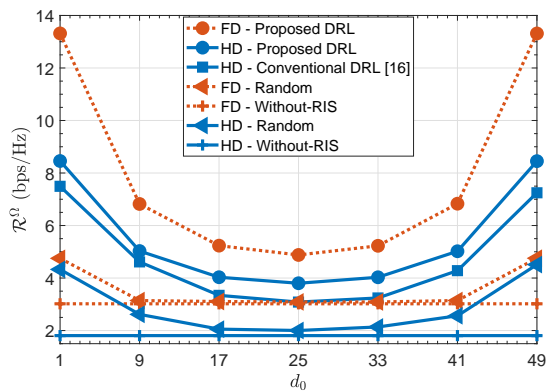}
    \caption{RIS deployment investigation.}
    \label{fig:varyd}
\end{figure}

{Figure.} \ref{fig:varyd} studies the impact of the RIS location on the system performance. 
It is shown that the proposed DRL algorithm significantly improves the rate for both operating modes, compared to the random phase shifts and without-RIS scenarios, especially when the RIS  {is located} closer to either the BS or {the} UE. On the other hand, the random phase shifts scenario does not improve the rate when the RIS is located relatively far from both BS and UE, compared to  {the scenario} without-RIS. Consequently, a proper optimization for the RIS phase shifts is needed to achieve a satisfactory performance. For the rest of the  {letter}, it is considered that $d_0 = \unit[1]{m}$.

\begin{figure}[h]
    \centering
    \includegraphics[scale= 0.4]{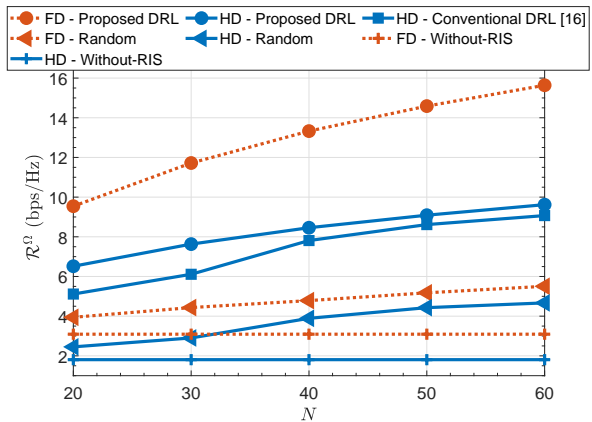}
    \caption{The impact of varying $N$ on the system performance.}
    \label{fig:varyN}
\end{figure}

{Figure.} \ref{fig:varyN} illustrates the effect of increasing $N$ on the system performance. As can be observed, $R^\Omega$ increases as $N$ increases for all algorithms. The proposed DRL algorithm provides an improvement of 4.6 bps/Hz and 8.5 bps/Hz in the achievable rate and sum-rate of the HD and FD modes, respectively, compared to the random phase shifts scenario at $N = 40$. It is worth noting that the gain gap increases as $N$ increases for the proposed DRL algorithm.

\begin{figure}[h!]
\centering
\begin{subfigure}{.34\textwidth}
  \centering
  \includegraphics[width=\linewidth]{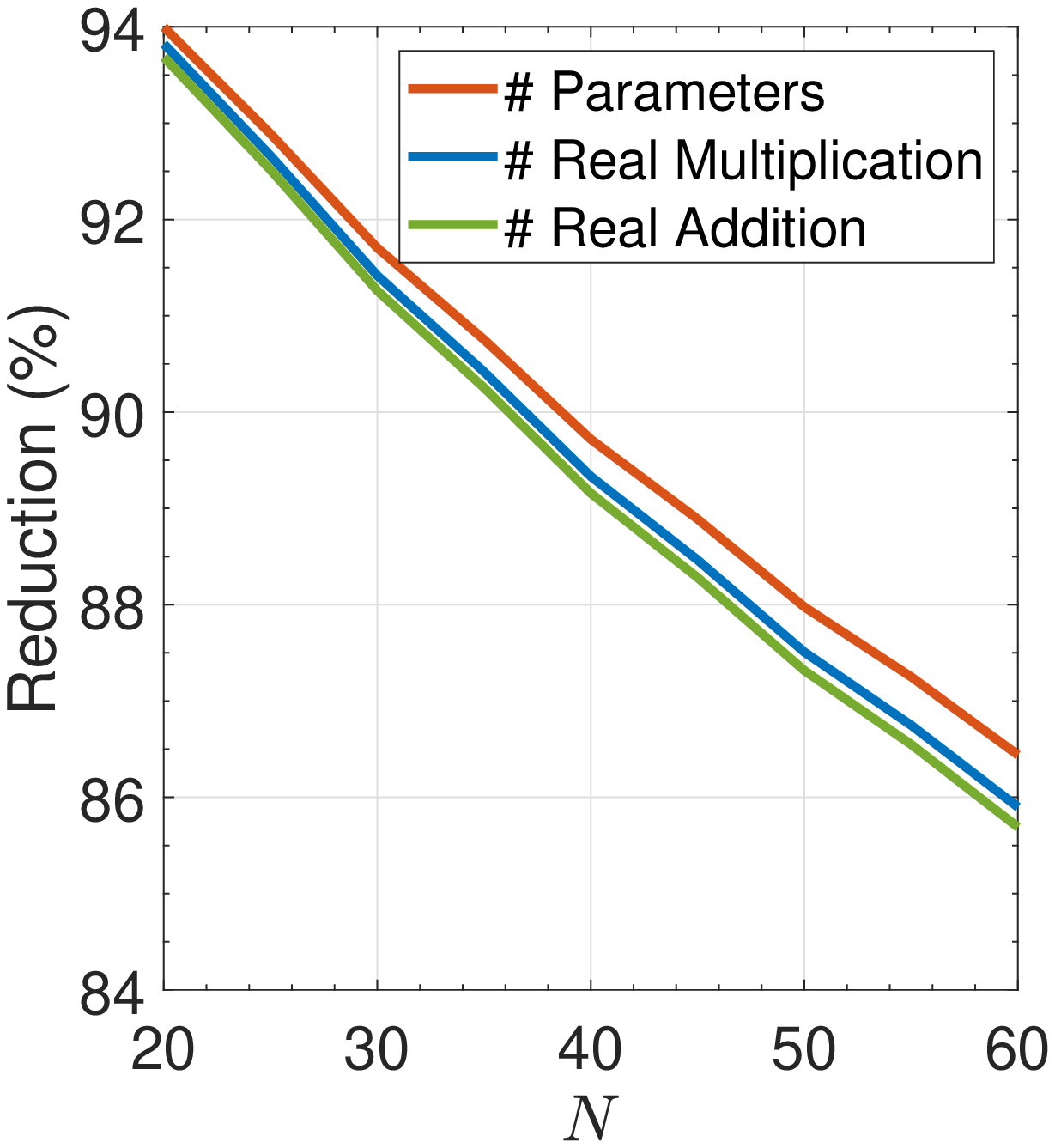}
  \caption{Practical range of $N$.}
  \label{fig:Comp_S}
\end{subfigure}%
\begin{subfigure}{.34\textwidth}
  \centering
  \includegraphics[width=\linewidth]{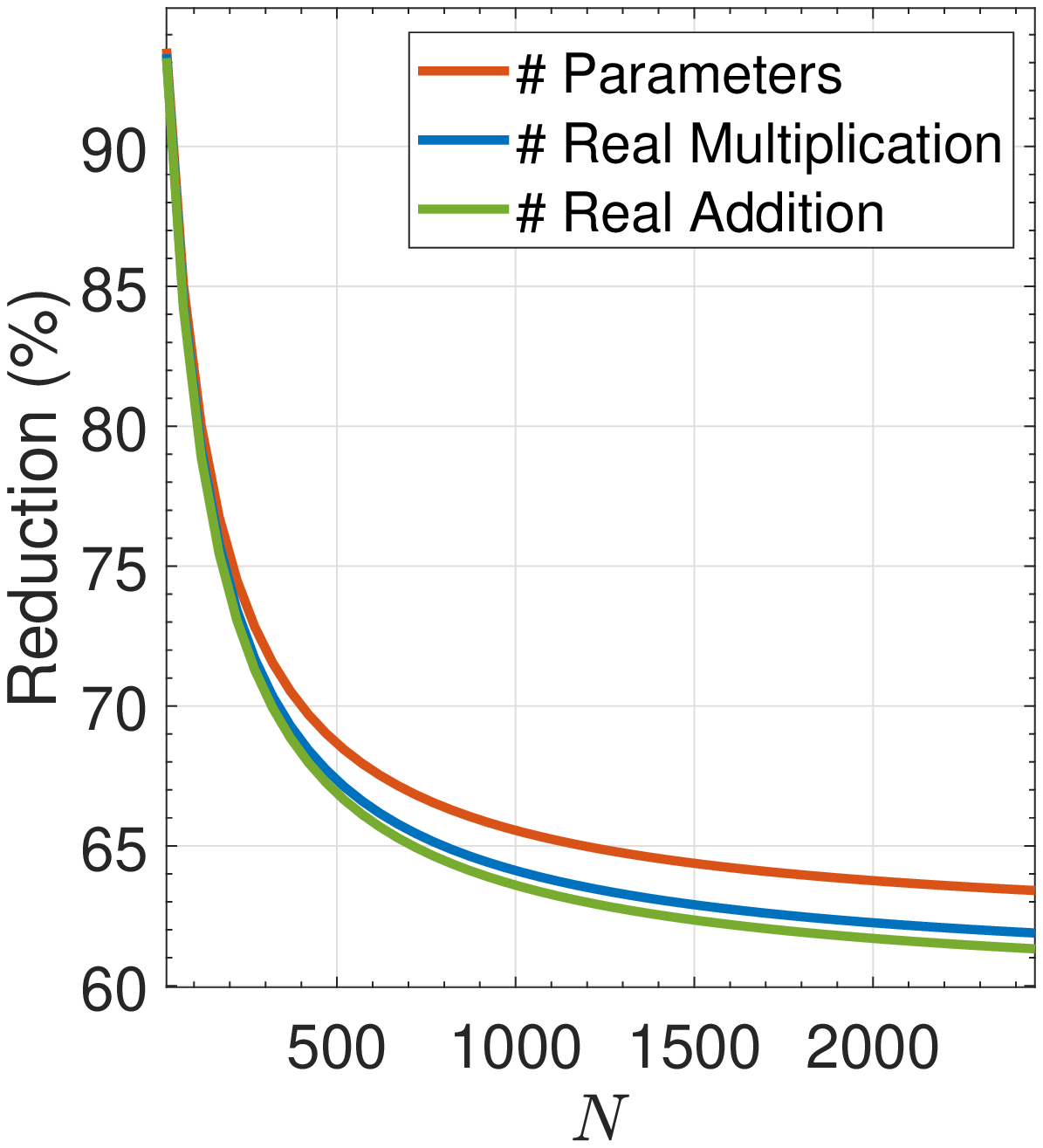}
  \caption{Asymptotic range of $N$.}
  \label{fig:Comp_L}
\end{subfigure}
\caption{Complexity reduction percentage versus $N$.}
\label{fig:Comp}
\end{figure}

In the HD operating mode, the proposed DRL algorithm improves the achievable rate performance by 1.4 bps/Hz and 0.6 bps/Hz at $N = 20$ and $N = 40$, respectively, when compared to \cite{DRLmain}, as depicted in Fig. \ref{fig:varyN}. Moreover, as shown in Fig. \ref{fig:Comp_S}, the proposed DRL algorithm in the HD mode (with $T = 800$ steps) significantly reduces the computational complexity of each NN in the range of 94\% to 86\% for the practical  {case} of $N = 20$ to $60$, respectively, compared to  {conventional} DRL in \cite{DRLmain} (with $T = 1000$ steps). Although the complexity reduction seems to decrease as $N$ increases, it saturates at 63\% for a certain large value of $N$, as seen from the asymptotic complexity bound in Fig. \ref{fig:Comp_L}.

Finally, the proposed DRL algorithm provides a significant improvement in the rate for both operating modes, compared with the random phase shifts scenario. Besides, with a 20\% {reduction in the number of} required steps {when compared with} the  {conventional} DRL algorithm, the proposed DRL algorithm  {guarantees a faster convergence and improves} the rate with {lower} computational complexity for each of the four NNs.

\section{Conclusion} \label{conclusion}
 {This letter considered} DRL for the rate maximization problem of the  {RIS-assisted} HD-FD MISO system,  {for the first time in the literature}.  {With a single} parameter setting, the proposed DRL algorithm  {optimized} the RIS phase shifts for both HD and FD operating modes. A novel DNN structure  {was} proposed to learn the optimal policy of the proposed DRL algorithm. Compared to the non-optimized scenario, the proposed DRL algorithm significantly  {improved} the rate for the HD and FD operating modes, respectively. Compared to the  {conventional} DRL algorithm in HD mode, the proposed DRL algorithm  {saved} 20\% of the required steps per episode  {and} {achieved} up to 1.4 bps/Hz rate improvement with up to 94\% reduction in the computational complexity.  {Future works can consider extending the proposed DRL algorithm to optimize the multi-user scenario.}


\vspace{-0mm}

\bibliographystyle{IEEEtran}
{\footnotesize\bibliography{IEEEabrv, mybibfile}}
\end{document}